\documentclass[reprint,aps,prd,nofootinbib,preprintnumbers,twocolumns,showpacs,10pt, superscriptaddress]{revtex4-1}
\usepackage{epsfig}
\usepackage{graphicx}
\usepackage{amsmath}
\usepackage{amsfonts}
\usepackage{amssymb}
\usepackage{bm}
\usepackage[usenames,dvipsnames]{color}
\usepackage{hyperref}
\hypersetup{
    colorlinks=true,       % false: boxed links; true: colored links
    linkcolor=NavyBlue,          % color of internal links (change box color with linkbordercolor)
    citecolor=Magenta}

\newcommand\ee{\end{equation}}
\newcommand\be{\begin{equation}}
\newcommand\eea{\end{eqnarray}}
\newcommand\bea{\begin{eqnarray}}

\newcommand\ie{{\it i.e.}~}
\newcommand\eg{{\it e.g.}~}

\newcommand\eq[1]{Eq.~(\ref{#1})}

\renewcommand\){\right)}
\renewcommand\[{\left[}
\renewcommand\]{\right]}

\begin{document}
\preprint{IFIC/14-37}

\def\thefootnote{\fnsymbol{footnote}}
\title{A model-independent fit to Planck and BICEP2 data}
\author{Laura Barranco} 
\affiliation{
Instituto de F\'isica Corpuscular (IFIC), CSIC-Universitat de Valencia,\\ 
Apartado de Correos 22085,  E-46071, Spain.}

\author{Lotfi Boubekeur}
\affiliation{ Instituto de F\'isica Corpuscular (IFIC), CSIC-Universitat de Valencia,\\
Apartado de Correos 22085,  E-46071, Spain.}
\affiliation{ Laboratoire de physique mathematique et subatomique (LPMS)\\
Universite de Constantine I, Constantine 25000, Algeria.}

\author{Olga Mena} 
\affiliation{
Instituto de F\'isica Corpuscular (IFIC), CSIC-Universitat de Valencia,\\ 
Apartado de Correos 22085,  E-46071, Spain.}

\begin{abstract}

{Inflation is the leading theory to describe elegantly the initial conditions that led to structure formation in our universe. 
In this paper, we present a novel phenomenological fit to the Planck, WMAP polarisation (WP) and the BICEP2 datasets using an alternative parameterisation. Instead of starting from inflationary potentials and computing the inflationary observables, we use a phenomenological parameterisation due to Mukhanov, describing inflation by an effective equation-of-state, in terms of the number of e-folds and two phenomenological parameters $\alpha$ and $\beta$. Within such a parametrisation, which captures the different inflationary models in a model-independent way, the values of the scalar spectral index $n_s$, its running and the tensor-to-scalar ratio $r$ are \emph{predicted}, given a set of parameters $(\alpha,\beta)$. We perform a Markov Chain Monte Carlo analysis of these parameters,  and we show that the combined analysis of Planck and WP data favours the Starobinsky and Higgs inflation scenarios.  Assuming that the BICEP2 signal is not entirely due to foregrounds, the addition of this last data set  prefers instead the $\phi^2$  chaotic models. The constraint we get from Planck and WP data alone on the \emph{derived} tensor-to-scalar ratio is $r<0.18$ at $95\%$~CL, value which is consistent with the one quoted from the BICEP2 collaboration analysis, $r = 0.16^{+0-06}_{-0.05}$, after foreground subtraction. This is not necessarily at odds with the $2\sigma$ tension found between Planck and BICEP2 measurements when analysing data in terms of the usual $n_s$ and $r$ parameters, given that the parameterisation used here includes, implicitly, a running spectral index.}

\end{abstract}

\maketitle

\twocolumngrid
\section{I. Introduction}
The recent claimed discovery of primordial $B$-modes by the BICEP2  collaboration \cite{bicep2,bicep22} has spurred a lot of interest in the cosmology community. One of the main topics of discussion is the tension between the BICEP2 results and the previous ones. In particular, this measurement corresponds to a tensor-to-scalar ratio of\footnote{This figure is obtained without subtracting polarised dust foregrounds, though the signal seen by BICEP2 outweigh any known foreground. Using the best available foreground template shifts the measured value to $r=0.16^{+0.06}_{-0.05}$.} $r=0.2^{+0.07}_{-0.05}$, while the Planck $TT$ data (combined with WP data, high-$\ell$ CMB measurements and without running of the scalar spectral index) \cite{planck,planck2} gives $r<0.11$ at $95\%$ CL. As argued in \cite{bicep2}, allowing for a running of the scalar spectral index makes the two datasets compatible at the one-sigma level. On the other hand, the large-field slow-roll models are able to explain successfully the BICEP2 data, however they predict negligible running, which, indeed, has not been seen in any previous observation like \eg Planck. This by itself suggests a non-trivial departure from the simple single-field slow-roll inflation paradigm. Plenty of effort has been devoted in the literature to reconcile BICEP2 and Planck observations, either by modifications of the inflationary sector~\cite{Contaldi:2014zua,Miranda:2014wga,Kawasaki:2014lqa,Abazajian:2014tqa} and/or of the standard cosmological scenario, as, for instance, extensions to the neutrino sector~\cite{Giusarma:2014zza,Zhang:2014dxk,Dvorkin:2014lea,Zhang:2014nta,Archidiacono:2014apa,DiValentino:2014zna}. Implications of the BICEP2 results in terms of the usual inflationary parameters  have also been extensively explored~\cite{Gerbino:2014eqa,Cheng:2014hba,Cheng:2014bma}.  In this work, we will look at this issue with a different perspective; we shall use an alternative parameterisation to fit both Planck and BICEP2 observations.
%\\
%\indent
There are two aspects of this discrepancy that are worth pursuing. The first one is purely experimental/observational and implies a re-assessing of all the systematic errors and possible unaccounted-for foregrounds (see \cite{Mortonson:2014bja} for a recent analysis in this direction). Despite the tremendous and impressive work done by the BICEP2 collaboration, this step is mandatory before drawing any definitive conclusion about the cosmological origin of this signal. For a roadmap of this program,  see \eg \cite{Dodelson:2014exa}. In the following, we will be assuming that the BICEP2 signal is primordial, although the novel phenomenological approach presented here can be applied to fit any cosmological data. The second aspect is theoretical, and it addresses the crucial question: did our universe suffer a quasi de Sitter expansion phase driven by the potential energy of a scalar field (the inflaton)? if yes, then, among the variety of available inflationary scenarios, which one describes better the observations? and, what are the physical implications of such a scenario?  In treating this last question, it is customary to use single-field slow-roll models as benchmark scenarios against which the temperature anisotropies observational data are tested. This is justified by the simplicity of these models when it comes to compute their predictions. Given a simple potential $V(\phi)$, where $\phi$ is the canonically-normalized inflaton field, one can compute easily the observational predictions in terms of the slow-roll parameters $\epsilon$ and $\eta$ defined as\footnote{For a nice review of slow-roll inflation see \eg  \cite{Lyth:2009zz}
. Throughout the paper, we will adopt natural units $\hbar=c=1$. As usual, the reduced Planck scale is given by $M_P=(8 \pi G_N)^{-1/2}\simeq 2.43\times 10^{18}$ GeV.} 
\be
\epsilon\equiv{1\over2} M_P^2\,\(V'/V\)^2 \textrm{ and } \eta\equiv M_P^2 \, V''/V ;  
\ee
where the primes denote derivatives with respect to $\phi$, \ie $V'\equiv d V/d\phi$ and so on. During slow-roll, these parameters are small \ie $\epsilon, |\eta|\ll1$, and the energy density of the universe is given approximately by the potential  $V\simeq 3 M_P^2 H^2$, where $H$ is the Hubble expansion rate during inflation. At leading order in slow-roll, the basic observables: the tensor-to-scalar ratio $r$ and the spectral index $n_s$, are given by 
\be
r=16\epsilon_\ast\textrm{ and } n_s= 1+2\eta_\ast -6 \epsilon_\ast \,,
\ee
where the subscript $\ast$ is to remind that quantities are evaluated at horizon exit.  These quantities are usually the basic ones used when testing models against observations. Each potential $ V(\phi)$ corresponds to a certain set of observables ($r$, $n_s$), but in general, these parameters are expected to be $O(1/N_\ast)$ where $N_\ast$ is the number of e-folds, starting from horizon exit, necessary to solve the standard cosmological problems. In general, this number has a mild dependence on the cosmological history, however under rather reasonable assumptions, $N_\ast$ takes values in the range $N_\ast\simeq 50$ - $60$, that we adopt from now on in our analysis.

Instead of the usual slow-roll parametrisation, one can use a more phenomenological and intuitive way of describing the inflationary phase through its equation of state \cite{Mukhanov:2013tua}. During inflation, the equation of state is $p\simeq -\rho\simeq - 3 H^2 M_P^2$, up to slow-roll corrections, while at the end of inflation $\dot\phi^2/2\simeq V(\phi)\simeq \rho/2$ and the equation of state is instead $p\simeq 0$. One can thus write that 
\be
{p\over \rho} = -1 + \frac{\beta}{(1+N_e)^\alpha}\, , 
\label{eq:ansatz}
\ee
where $\alpha$ and $\beta$ are phenomenological parameters and are both positive and of $O(1)$, and $N_e$ is the number of remaining e-folds to end inflation $N_e(t)\equiv\int^{t_e}_t dt\, H$ and it runs from $N_\ast$,  at horizon exit, to 0, when inflation ends. Using energy conservation $\dot\rho+3H (\rho+p)=0$ one gets the following expressions for the tilt and tensor fraction  \cite{Mukhanov:2013tua}
\begin{subequations}
\label{eq:nsr}
\bea
n_s-1&=&-3\frac{\beta}{\(N_\ast+1\)^\alpha}-\frac{\alpha}{N_\ast+1}
\label{eq:ns}\\
r&=&24\beta\over \(N_\ast+1\)^\alpha
\label{eq:r}
\eea
\end{subequations}
\begin{figure}[!t]
\begin{center}
\hspace{-1cm}
\includegraphics[width=0.4\textwidth]{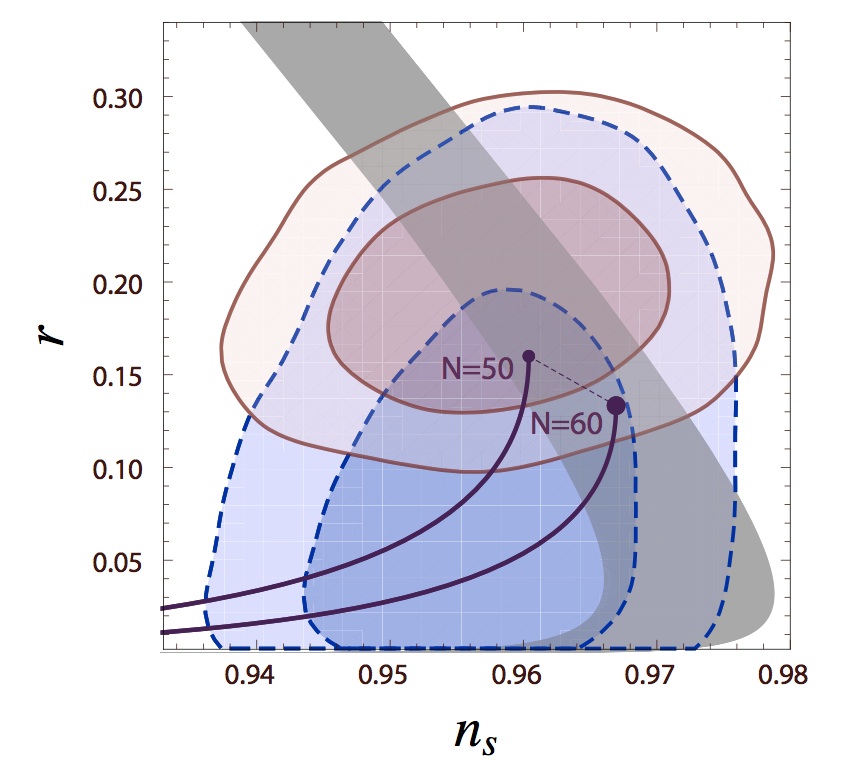} 
\caption{Confidence regions in the $(n_s, r)$ parameterisation plane. The blue (dashed boundary) areas represent the $68\%$ and $95\%$ CL regions of Planck (including a non-zero running), while the red (solid boundary) areas are the $68\%$ and $95\%$ CL regions for BICEP2 only. The grey band represents the predictions of the models captured by the parameterisation \eq{eq:ansatz} for $50\le N_\ast\le 60$. The solid  magenta lines correspond to the natural inflation scenario. For large decay constant $f\gg M_P$, they reduce to the $V\propto\phi^2$ scenario (short-dashed magenta line).}
\label{fig:fig1}
\end{center}
\end{figure}

The general prediction of the ansatz Eq.~(\ref{eq:ansatz}) is that the tilt is always {\it negative}, irrespective of the inflationary scenario. In contrast, the value of the tensor-to-scalar ratio can take any value depending on both $\alpha$ and $\beta$. In addition, one can compute the running of the tilt 
\be
\alpha_s\equiv d n_s/d \log k = -{3\alpha\beta \over (1+N_\ast)^{\alpha +1}}-{\alpha\over (N_\ast+1)^2}\, , 
\label{eq:run}
\ee
which is, like the tilt, always {\it negative}.

The parameterisation Eq.~(\ref{eq:ansatz}) encodes a variety of models with completely different predictions \cite{Mukhanov:2013tua}. Notice however that this phenomenological description of the inflationary phase is not  completely equivalent to the slow-roll picture, as there is no more freedom in the signs of both the tilt and the running.

From Fig.~\ref{fig:fig1},   it is clear that  the observationally preferred value of the scalar spectral index $n_s\simeq 0.96$ corresponds to two different branches. The first one lies close to the horizontal line  $r\approx 0$, in Fig.~\ref{fig:fig1}, and contains for instance Starobinsky models of inflation \cite{Starobinsky:1980te}, which are based on the Lagrangian $\sqrt{-g}\(R+a R^2\)$. In terms of the phenomenological parameterisation Eq.~(\ref{eq:ansatz}), this branch corresponds to $\alpha=2$, and $r\simeq 10^{-2}\beta$. In particular \cite{Mukhanov:2013tua}, Starobinsky inflation corresponds to $\beta=1/2$. 

In contrast, the second branch, with significantly higher tensor fraction (appearing as a thick diagonal grey area in Fig.~\ref{fig:fig1}) is where chaotic inflation models, $V(\phi)\propto\phi^n$~\cite{Linde:1983gd}, live. In terms of the parameterisation \eq{eq:ansatz}, chaotic scenarios live on the line corresponding to $\alpha=1$. From Eqs.~(\ref{eq:nsr}), the line in the $(n_s, r)$ plane is given by $n_s\simeq 1-\frac{r}8$, up to a $O(1/N_\ast)$ correction. On the other hand, the parameter $\beta$ fixes the power of the potential $V\propto\phi^n$, as  $n=6\beta$.

\begin{figure}[!t]
\begin{center}
\hspace{-1.5cm}
\includegraphics[width=0.42\textwidth]{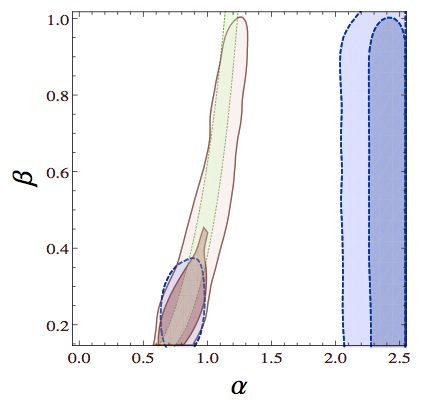} 
\caption{Confidence regions in the $(\alpha, \beta)$ parameters in Eq.~(\ref{eq:ansatz}). The red areas (solid  boundary) represent the $68\%$ and $95\%$ CL allowed regions arising from a combined analysis of the Planck, WP and BICEP2 data, while the blue areas (dashed boundary) are the $68\%$ and $95\%$ CL allowed regions from the analysis of Planck and WP data. The green region with dotted contours represent the joint 1$\sigma$  preferred region for Planck and BICEP2.}
\label{fig:25a}
\end{center}
\end{figure}

The natural inflation scenario \cite{Freese:1990rb, Adams:1992bn}, in which the inflaton is a pseudo Nambu-Goldstone boson (pNGB), is represented by the purple line in Fig.~\ref{fig:fig1}. This scenario, described by the potential $V(\phi)\propto \[1-\cos(\phi/f)\]$, is captured by the paramterisation  Eq.~(\ref{eq:ansatz}) but only for large enough decay constants $f\gtrsim 10 M_P$. We recall that in the limit of very large decay constant, $f\gg M_P$, Natural inflation reduces to the $\phi^2$ scenario represented 
by the thick purple dots ($N_\ast=50$ and $N_\ast=60$) in  Fig.~\ref{fig:fig1}.

Before describing our cosmological data fits, let us determine the interval spanned by the phenomenological parameters $\alpha$ and $\beta$. First, as explained in Ref.~\cite{Mukhanov:2013tua}, given that inflation ends, \ie $N_e\to 0$, when  $p/\rho\approx 0$, it follows that  $\beta$ cannot be much larger that 1. 
Second, given that in the most optimistic situation, the tensor-to-scalar ratio will be measured at an accuracy of \cite{Dodelson:2014exa} $\Delta r/r=10^{-2}$, it is clear from \eq{eq:r} that\footnote{A meaningful measurement of the tensor-to-scalar ratio implies that $\Delta r \lesssim r$. Using \eq{eq:r}, one gets
$$
\alpha\lesssim {\log (24 \beta/\Delta r)}/{\log(N_\ast+1)}\, , 
$$
which for $\beta\lesssim 1$ and the optimistic percent-level observational error on $r\simeq 0.001$ targeted \eg by {\sl COrE} \cite{CORE} and {\sl PIXIE} \cite{pixie} gives $\alpha\lesssim 2.5$. 
 Notice that the above estimate does not change appreciably as it depends only logarithmically on both $N_\ast$ and $\Delta r$. } $\alpha\lesssim 2.5$. We shall adopt these priors  in our numerical analyses.

The structure of the paper is as follows. In section~\ref{sec:sec2}, we describe the method followed when performing the  fits to the different datasets. Next, in Section \ref{sec:sec3}, we present our results in terms of the parameters $\alpha$ and $\beta$ governing the parameterisation \eq{eq:ansatz},  and in terms of the \emph{derived}, most commonly used inflationary  parameters $n_s$ and $r$. We also discuss their implications.  Finally, we draw our conclusions in Sec.~\ref{sec:sec4}.

\begin{table}[t!]
\begin{center}
\begin{tabular}{c|c}
\hline\hline
 Parameter & Prior\\
\hline
$\Omega_{b}h^2$ & $0.005 \to 0.1$\\
$\Omega_{c}h^2$ & $0.001 \to 0.99$\\
$\Theta_s$ & $0.5 \to 10$\\
$\tau$ & $0.01 \to 0.8$\\
$\log{(10^{10} A_{s})}$ & $2.7 \to 4$\\
$\alpha$ & $0 \to 2.5$\\
 $\beta$&  $0 \to 1$\\
\hline\hline
\end{tabular}
\caption{Uniform priors for the cosmological parameters considered here.}
\label{tab:tabpriors}
\end{center}
\end{table}

\section{Data analysis}
\label{sec:sec2}
\subsection{Method}

The phenomenological scenario we explore is described by the following parameter set:
\begin{equation}
\label{parameter}
  \{\omega_b, \omega_c, \Theta_s, \tau, \log[10^{10}A_{s}],  \alpha, \beta\}~,
\end{equation}

\noindent where $\omega_b\equiv\Omega_bh^{2}$ and $\omega_c\equiv\Omega_ch^{2}$ are
the physical baryon and cold dark matter energy densities,
$\Theta_{s}$ is the ratio between the sound horizon and the angular
diameter distance at decoupling, $\tau$ is the reionization optical depth, 
$A_{s}$ the amplitude of the primordial spectrum and $\alpha$ and $\beta$ are the phenomenological parameters 
governing the parameterisation given in Eq.~(\ref{eq:ansatz}). In the following, we fix the number of e-folds to $N_\ast=60$.  Furthermore, we assume that dark energy is described by a cosmological constant. Table~\ref{tab:tabpriors} specifies the priors considered on the cosmological parameters listed above. The commonly used $(n_s,r)$ parameters can be easily obtained using Eqs.~(\ref{eq:nsr}), however unlike the usual case where the running of the spectral index is a free parameter, the running here is completely fixed through \eq{eq:run}, given $(\alpha, \beta)$ and $N_\ast$. In our analysis, we also consider the so-called inflation consistency relation, relating the tensor spectral index to $r$ through $n_T = -r/8$, which is also valid in this parametrisation\footnote{See Eq.~(8.125) of \cite{Mukhanov:2005sc}.}. For our numerical calculations, we use the CAMB Boltzmann code~\cite{camb}, deriving posterior distributions for the cosmological parameters from the datasets described in the next section by means of Monte Carlo Markov Chain (MCMC) analyses. Our MCMC results rely on the publicly available MCMC package \texttt{cosmomc}~\cite{Lewis:2002ah} that implements the Metropolis-Hastings algorithm. 

\subsection{Cosmological data}

In our analyses we will consider, as a basic dataset: the  Planck CMB temperature anisotropies data \cite{Ade:2013ktc,Planck:2013kta} together with the 9-year polarization data from the WMAP satellite~\cite{Bennett:2012fp}.   
The total likelihood for the former data is obtained by means of  the Planck collaboration publicly available likelihood code, see Ref.~\cite{Planck:2013kta} for details. The Planck temperature power spectra extend up to a maximum multipole  number $\ell_{\rm max}=2500$,  while the WMAP 9-year polarization data (WP) is analysed up to a maximum multipole $\ell=23$~\cite{Bennett:2012fp}.

As stated before, very recently, the BICEP2 collaboration has found evidence for the detection of $B$-modes in the multipole range
$30 <\ell< 150$ spanned by their three-year dataset~\cite{bicep2,bicep22}, with $6\sigma$ significance. The detected $B$-mode signal exceeds any known systematics and/or expected foregrounds and is well fitted with a tensor-to-scalar ratio $r=0.2^{+0.07}_{-0.05}$. The BICEP2 likelihood has been properly accounted for in our MCMC numerical analyses, by using the latest version of \texttt{cosmomc}.

\begin{figure}[!t]
\begin{center}
\hspace{-1cm}
\includegraphics[width=0.5\textwidth]{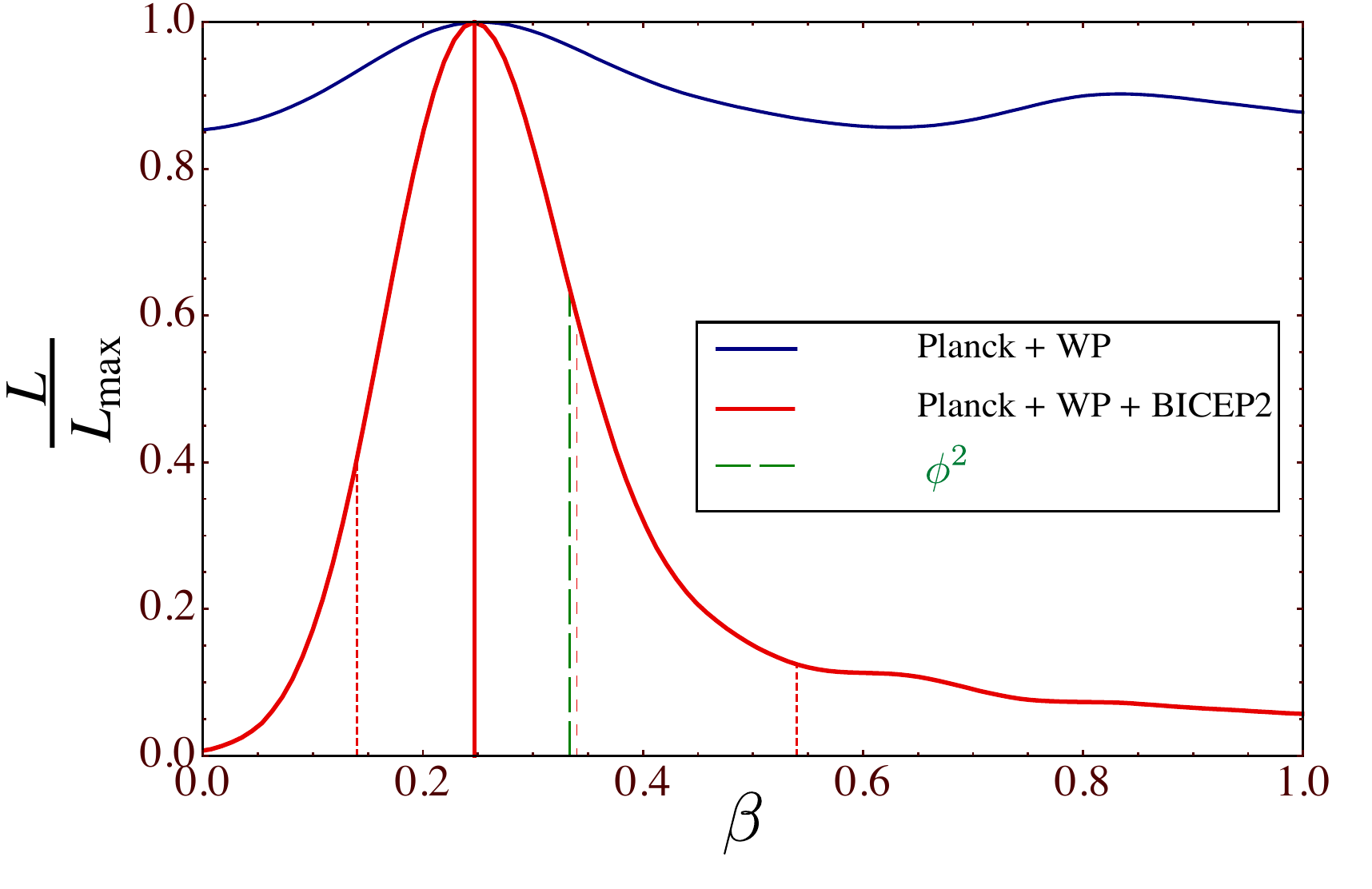} 
\caption{The derived likelihood distribution for the phenomenological  parameter $\beta$ of \eq{eq:ansatz} using different datasets. The red thick (solid and dashed) vertical lines represent the best-fit ($\pm 1 \sigma$ intervals) of $\beta$, while the red thin-dashed  line stands for the derived mean value of $\beta$, see Table \ref{tab:table25} for details. The quadratic chaotic scenario, corresponding to $\beta=1/3$, is represented by a green long-dashed line.}
\label{fig:beta1d}
\end{center}
\end{figure}

\begin{figure}[!t]
\begin{center}
%\hspace{-2cm}
\includegraphics[width=0.4\textwidth]{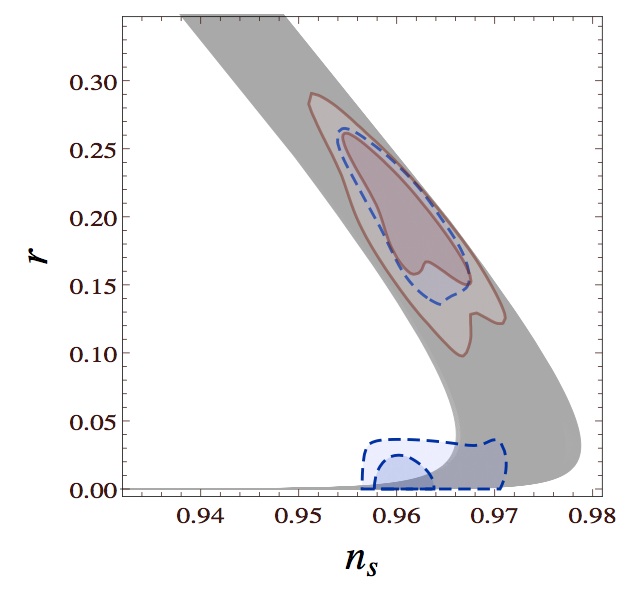} 
\caption{Confidence regions of the \emph{derived} parameters $(n_s, r)$ using the parameterisation given by Eq.~(\ref{eq:ansatz}). The colour coding is the same than in Fig.~1. The grey band represents the predictions of the models covered by the parameterisation given by Eq.~(\ref{eq:ansatz}) for $50\le N_\ast\le60$.
}
\label{fig:25ns}
\end{center}
\end{figure}

\section{Results}

\label{sec:sec3}
We represent the results of our MCMC analyses both in the $(\alpha, \beta)$ plane and in the usual $(n_s, r)$ plane. Figure~\ref{fig:25a} shows the $68\%$ and $95\%$ CL contours in $(\alpha, \beta)$. The red solid contours depict the $68\%$ and $95\%$ CL allowed regions from the combined analysis of  Planck, WP and BICEP2 data, while the blue dashed contours refer to the $68\%$ and $95\%$ CL allowed regions from the analysis of Planck and WP data. The green dotted region represents the limits in the ($\alpha$, $\beta$) plane inferred from the $1\sigma$ preferred values for $n_s$ and $r$ from Planck and BICEP2 data, respectively. Notice that the results from our MCMC analyses after the combination of Planck, WP and BICEP2 datasets lie precisely within this region. The combination of Planck and WP data is completely insensitive to the $\beta$ parameter, as $\beta$ sets the amount of gravitational waves. The addition of BICEP2 data, however, strongly constrains the value of $\beta$, as illustrated in Fig.~\ref{fig:beta1d}, which shows the one-dimensional probability density for the $\beta$ parameter before and after the inclusion of BICEP2 measurements.  Figure~\ref{fig:beta1d} shows as well the best-fit and the $1\sigma$ allowed regions for the $\beta$ parameter after considering all the measurements exploited in this study. We also depict in Fig.~\ref{fig:beta1d} the value of $\beta$ for the most favoured inflationary scenario, as we shall see in what follows.

Figure~\ref{fig:25ns} depicts the $68\%$ and $95\%$ CL allowed contours in the plane of the \emph{derived} parameters $n_s$ and $r$, together with the region covered by the parameterisation given by Eq.~(\ref{eq:ansatz}) for $50\le N_\ast\le 60$. Table~\ref{tab:table25} shows the constraints at 68\% confidence level on the cosmological parameters considered in our MCMC analyses for the different data combinations explored here. Notice that, when BICEP2 measurements are not considered, $\alpha=2.24 \pm 0.43$ while $\beta=0.50\pm 0.28$, which corresponds to $n_s\simeq 0.96$ and $r<0.18$ at $95\%$~CL, values that can clearly be inferred from the results depicted in  Fig.~\ref{fig:25ns}. The constraint we get for Planck and WP data alone is $r<0.18$ at $95\%$~CL, value to be compared with the value quoted for BICEP2 collaboration for $r = 0.16^{+0.06}_{-0.05}$~\cite{bicep2} after subtracting the various foregrounds. Therefore, the upper limits we get on the tensor-to-scalar ratio $r$ from Planck and WP data using the parameterisation given in Eq.~(\ref{eq:ansatz}) are very close to the figure of  $r = 0.16^{+0.06}_{-0.05}$ reported by the BICEP2 collaboration. This is consistent with the tension found between BICEP2 and Planck using the  standard parameters $n_s$ and $r$, as the parameterisation used here includes implicitly a non-vanishing running spectral index, see Eq.~(\ref{eq:run}).

The resulting favoured values of $n_s\simeq 0.96$ and $r\approx 0$ from the Planck and WP data analysis may be associated to the Starobinsky model of inflation \cite{Starobinsky:1980te}. Indeed, in terms of the phenomenological parameterisation \eq{eq:ansatz}, Starobinsky inflation corresponds to $\alpha= 2$ and $\beta=1/2$~\cite{Mukhanov:2013tua}. Another inflationary scenario that can also be identified with these values of $\alpha$ and $\beta$ is Higgs inflation, in which the Standard Model Higgs boson itself is responsible for inflation~\cite{Bezrukov:2007ep,Bezrukov:2013fka}. Higgs inflation predicts a scalar spectral index $n_s\simeq 0.97$ and a tensor-to-scalar ratio $r\simeq 0.0033$ for $N_\ast=60$~\cite{Bezrukov:2007ep} and is indistinguishable observationally from the Starobinsky model.

\begin{figure}[!t]
\begin{center}
\hspace{-1.5cm}
\includegraphics[width=0.5\textwidth]{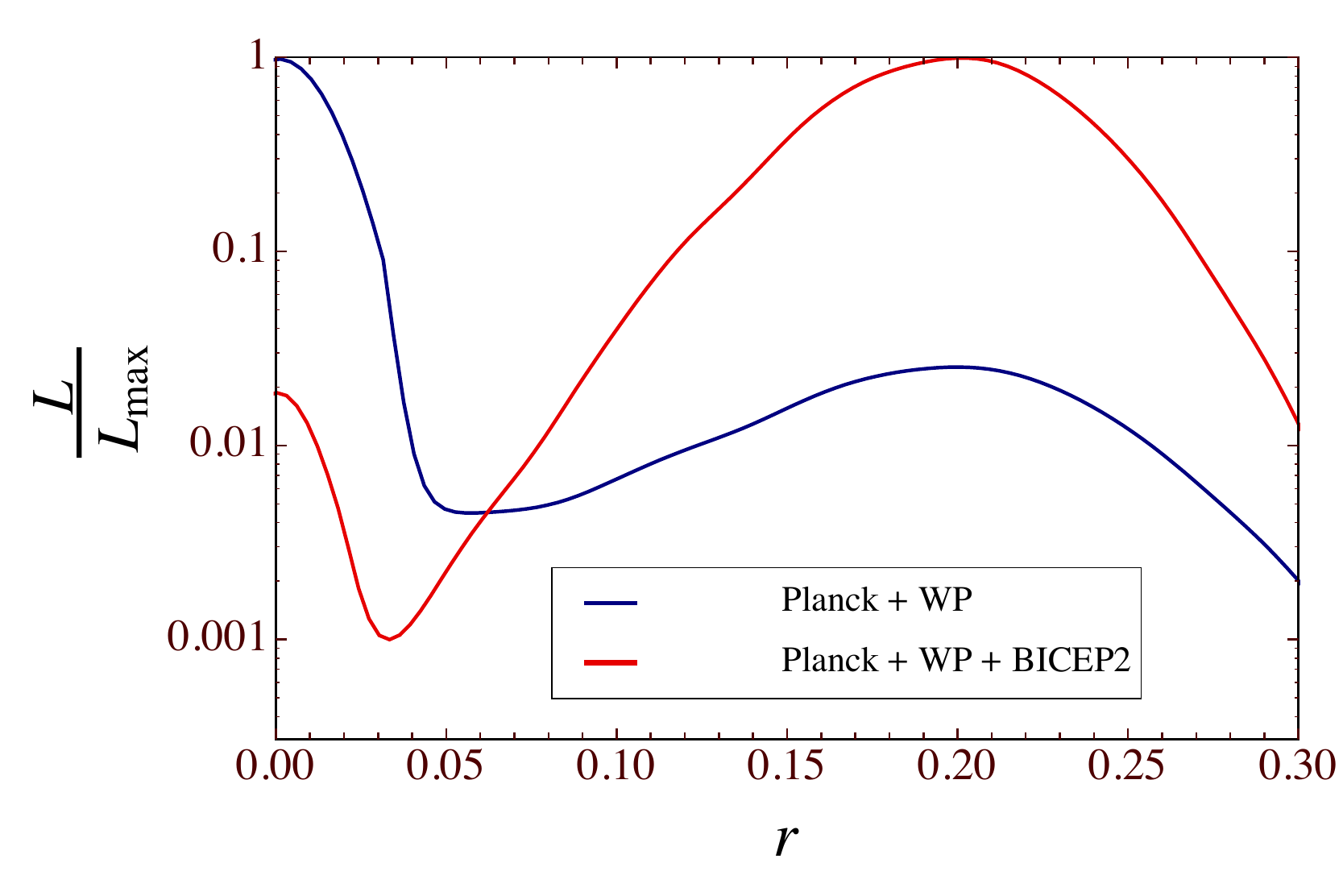} 
\caption{The derived posterior likelihood distribution for the tensor-to-scalar ratio using different datasets. }
\label{fig:like}
\end{center}
\end{figure}

We can learn from the red contours in Figs.~\ref{fig:25a} and \ref{fig:25ns} that, when adding to Planck and WP data the BICEP2 mesurements, models with such large values of $\alpha \sim 2$ are no longer favoured. The resulting mean values of the two parameters are $\alpha=0.88 \pm 0.17$ and $\beta=0.34\pm 0.20$, which correspond to $n_s= 0.961 \pm 0.003$ and $r=0.195 \pm 0.037$ (see Table \ref{tab:table25}). This value of $r$ belongs to the second of the branches associated to $n_s \simeq 0.96$, which are depicted by the thick diagonal grey area in Figs.~\ref{fig:fig1} and \ref{fig:25ns}.  The one-dimensional posterior probability densities for the \emph{derived} scalar-to-tensor ratio $r$ are depicted in Fig.~\ref{fig:like}. Notice that for the two possible data combinations the probability distribution is bimodal, showing two maxima: one is located at $r\simeq 0$ and the other one is located at $r\simeq 0.2$. These two peaks stand for the two possible values of $r$ corresponding to $n_s \simeq 0.96$. Each of them is located in one of the two branches shown in Figs.~\ref{fig:fig1} and \ref{fig:25ns}. While the probability distribution function for Planck + WP data has a global maximum at the $r\simeq 0$ branch, the addition of BICEP2 measurements displaces the global maximum towards the $r\simeq 0.2$ region in the other possible branch. As explained before, it is precisely in this second branch where chaotic inflation models,  $V(\phi)\propto\phi^n$, live~\cite{Mukhanov:2013tua}. 
Chaotic models with quadratic (quartic) potentials predict $n_s\simeq 0.96$ and $r\simeq 0.16$ ($n_s\simeq 0.94$ and $r\simeq 0.32$)~\cite{Mukhanov:2013tua}. Therefore, the mean values of $\alpha$ and $\beta$ resulting from the combined analyses of Planck, WP and BICEP2 data seem to favour $\phi^2$ models of chaotic inflation and highly disfavour Starobinsky and Higgs inflation scenarios. The quartic chaotic model also is disfavoured with respect to the quadratic one. The status of the former two inflationary models has also been explored recently in the literature (see \eg Refs.~\cite{Ferrara:2014ima,Cook:2014dga,Nakayama:2014koa}) where it has been found that these two models require either extreme fine-tuning or non-trivial extensions to be compatible with BICEP2 results. Chaotic inflationary models have also been recently revisited in a number of analyses~\cite{Creminelli:2014oaa,Okada:2014lxa,DiBari:2014oja,Kaloper:2014zba,Creminelli:2014fca}. On the other hand, Natural inflation is the only case which can not be analysed in terms of $\alpha$ and $\beta$ except in the large decay constant regime \ie $f\gg M_P$, where $M_P$ is the Planck mass. In this case, the constraints are similar to the case of the quadratic chaotic scenario (see Fig.~\ref{fig:fig1}). Our derived bound on the tensor-to-scalar ratio ($r<0.18$ at 95$\%$ CL)  does not put significant constraints on $f$.  If on the other hand, we include the BICEP2 datasets, we can translate the 1$\sigma$ interval into a lower bound on the decay constant $f\gtrsim 44.72 M_P$ for $N_\ast=50$. This makes Natural inflation practically indistinguishable from the quadratic chaotic scenario, given the present precision. The next generation of observations will improve the situation considerably, allowing for instance to distinguish between the two scenarios if $f\lesssim 30 M_P$. See \eg Refs.~\cite{Creminelli:2014oaa,Freese:2014nla,Kohri:2014rja} for recent appraisals of the Natural inflation scenario. The results previously discussed have been obtained fixing the number of e-folds to $N_\ast=60$. Assuming $N_\ast=50$ instead does not change the main conclusions outlined above.

\begin{table*}[!t]
\begin{center}
\begin{tabular}{|c|c|c|c|c|c|}
\hline\hline
Parameter & Planck+WP & Planck+WP+BICEP2\\
\hline
$\Omega_bh^2$ &$0.0209\pm0.0002$ & $0.0209\pm0.0002$ \\
$\Omega_ch^2$ &$0.1165\pm0.0018$ & $0.1167\pm0.0020$\\
$\theta$ &$1.0409\pm0.00055$ & $1.0408\pm0.00055$ \\
$\tau$ &$0.086\pm0.015$ & $0.078\pm0.013$ \\
$\log[10^{10} A_s]$ &$3.063\pm0.031$ & $3.047\pm0.026$\\
$\alpha$ &$2.24 \pm 0.43$&$0.88\pm 0.17$ \\
$\beta$ &$0.50\pm 0.28 $&$0.34\pm 0.20$ \\
$n_s$\emph{(derived)} &$0.961\pm 0.002$&$0.961\pm 0.003$\\
$r$ \emph{(derived)} &$<0.18$&$0.195 \pm 0.037$ \\

\hline\hline
 \end{tabular}
 \caption{Constraints at 68\% confidence level on cosmological parameters from our analyses for Planck+WP and Planck+WP+BICEP2 data. When quoting upper bounds, we show the $95\%$ CL limits. Notice that the scalar spectral index and the tensor to scalar ratio are \emph{derived} parameters.}
 \label{tab:table25}
 \end{center}
 \end{table*}

We conclude this section commenting on the results obtained when using a slightly different upper prior on the $\alpha$ parameter. In general, smaller values of $\alpha$ will give rise to a higher tensor-to-scalar ratio and therefore the tension between Planck and BICEP2 measurements may be alleviated. 
If we assume an upper prior on $\alpha$ of $2$,  the $95\%$~CL upper bound on the derived tensor-to-scalar ratio parameter is slightly larger ($r<0.23$ at $95\%$~CL). On the contrary, when higher values for $\alpha$ are considered, the significance of the tension between Planck and BICEP2 measurements slightly increases, as higher values of $\alpha$ correspond to lower values of $r$. A fit to Planck and WP data gives an upper limit on $r<0.17$ at $95\%$~CL when using an upper prior on $\alpha$ of $3$. When BICEP2 measurements are included in the analysis, we obtain $n_s=0.961\pm 0.004$ and $r=0.184 \pm 0.040$ ($n_s=0.961\pm 0.004$ and $r=0.192 \pm 0.037$) for an upper prior on $\alpha$ of $2$  ($3$).  These results  are almost identical to the ones quoted in Table~\ref{tab:table25}.  We have also checked that the posterior  probability density profiles for both the parameter $\beta$ and the tensor-to-scalar ratio $r$ do not exhibit a significant prior dependence. Summarising, the effect of the upper prior choice on $\alpha$ barely changes our main results.

\section{Conclusions}
\label{sec:sec4}

The recent claimed discovery of primordial gravitational waves by the BICEP2 collaboration has opened a new window into the inflationary paradigm. Chaotic inflation scenarios, highly disfavoured by Planck temperature data, are, after BICEP2 results, among the most plausible ones.  Model-independent data analyses are usually presented in terms of the scalar spectral index $n_s$ and the tensor-to-scalar ratio $r$, which can then be related to a particular model via the inflationary slow-roll parameters. Here we employ an alternative parameterisation due to Mukhanov, describing inflation by an effective equation of state, which captures most of the relevant inflationary scenarios (at least in their basic formulation). Using this parameterisation, one can easily identify the different models as well as \emph{derive} the usual $n_s$ and $r$ parameters.  The effective equation of state used here is described by only two parameters, $\alpha$ and $\beta$, since the running of the spectral index $\alpha_s$ is no longer a free parameter, as is unambiguously determined once the values of $\alpha$ and $\beta$ are fixed.

Using Markov Chain Monte Carlo methods, we show that the combined analyses of Planck temperature and WMAP polarisation (WP) data are unable to determine $\beta$, as this last parameter sets the amount of gravitational waves through \eq{eq:r}. However, these two datasets are able to constrain the other parameter involved, $\alpha$, resulting in a mean value $\alpha=2.24 \pm 0.43$, which corresponds to $n_s \simeq  0.96$. Such value of $\alpha$, favoured by the Planck and WP data analyses is associated to both Starobinsky and Higgs inflationary models. The constraint we get on the \emph{derived} tensor-to-scalar ratio, $r<0.18$ at $95\%$~CL, is perfectly consistent with the value quoted from the BICEP2 collaboration ($r = 0.16^{+0.06}_{-0.05}$~\cite{bicep2}) after subtracting the various foregrounds. However, this is not necessarily in conflict with the $2\sigma$ tension found between Planck and BICEP2 measurements when analysing data in terms of the usual $n_s$ and $r$ parameters, since the parameterisation used here includes a running which depends exclusively on the $\alpha$ and $\beta$ parameters, as well as on the number of e-folds. 

The addition of BICEP2 data to Planck and WP measurements strongly constrains the values of the phenomenological parameters to the values $\beta\simeq 1/3$, and $\alpha \simeq 1$. Such values of $\alpha$ correspond to chaotic inflationary models, characterised by a potential $\phi^n$, where  $n=6\beta$. Therefore, the results from the combined analysis of Planck, WP and BICEP2 data strongly favour $\phi^2$ models of chaotic inflation and rule-out Starobinsky and Higgs inflation scenarios. Upcoming polarisation data from Planck may confirm or falsify the $\phi^2$ scenario as the most plausible one for the inflationary period. Future CMB missions, such as  {\sl COrE} \cite{CORE} and {\sl PIXIE} \cite{pixie}, combined with galaxy clustering and weak lensing data from the Euclid survey~\cite{Amendola:2012ys} hold the key to establish the amount of primordial $B$-modes and the ensuing  theoretical implications, especially if the tensor-to-scalar ratio is as large as suggested by BICEP2.

\acknowledgments
We are grateful to Paolo Creminelli, Scott Dodelson, Carlos Pe\~na Garay and Aaron C. Vincent for their unvaluable comments and suggestions on this work.  O.M. is supported by the Consolider Ingenio project CSD2007-00060, by PROMETEO/2009/116, by the Spanish Ministry Science project FPA2011-29678 and by the ITN Invisibles PITN-GA-2011-289442.


\begin{thebibliography}{99}

\twocolumngrid
  

  
  \bibitem{bicep2}
  P.~A.~R.~Ade {\it et al.}  [BICEP2 Collaboration],
 %``BICEP2 I: Detection Of B-mode Polarization at Degree Angular Scales,'' 
 arXiv:1403.3985 [astro-ph.CO].
  \bibitem{bicep22} 
  P.~A.~R.~Ade {\it et al.}  [BICEP2 Collaboration],
  %``BICEP2 II: Experiment and Three-Year Data Set,''
  arXiv:1403.4302 [astro-ph.CO].


  
  \bibitem{planck}
 P.~A.~R.~Ade {\it et al.}  [Planck Collaboration],
  %``Planck 2013 results. XVI. Cosmological parameters,''
  arXiv:1303.5076 [astro-ph.CO].

\bibitem{planck2}  
P.~A.~R.~Ade {\it et al.}  [Planck Collaboration],
  %``Planck 2013 results. XXII. Constraints on inflation,''
  arXiv:1303.5082 [astro-ph.CO].  


\bibitem{Contaldi:2014zua} 
  C.~R.~Contaldi, M.~Peloso and L.~Sorbo,
  %``Suppressing the impact of a high tensor-to-scalar ratio on the temperature anisotropies,''
  arXiv:1403.4596 [astro-ph.CO].


\bibitem{Miranda:2014wga} 
  V.~Miranda, W.~Hu and P.~Adshead,
  %``Steps to Reconcile Inflationary Tensor and Scalar Spectra,''
  arXiv:1403.5231 [astro-ph.CO].


\bibitem{Kawasaki:2014lqa} 
  M.~Kawasaki and S.~Yokoyama,
  %``Compensation for large tensor modes with iso-curvature perturbations in CMB anisotropies,''
  arXiv:1403.5823 [astro-ph.CO].


\bibitem{Abazajian:2014tqa} 
  K.~N.~Abazajian, G.~Aslanyan, R.~Easther and L.~C.~Price,
  %``The Knotted Sky II: Does BICEP2 require a nontrivial primordial power spectrum?,''
  arXiv:1403.5922 [astro-ph.CO].


\bibitem{Giusarma:2014zza} 
  E.~Giusarma, E.~Di Valentino, M.~Lattanzi, A.~Melchiorri and O.~Mena,
  %``Relic Neutrinos, thermal axions and cosmology in early 2014,''
  arXiv:1403.4852 [astro-ph.CO].


\bibitem{Zhang:2014dxk} 
  J.~-F.~Zhang, Y.~-H.~Li and X.~Zhang,
  %``Sterile neutrinos help reconcile the observational results of primordial gravitational waves from Planck and BICEP2,''
  arXiv:1403.7028 [astro-ph.CO].


\bibitem{Dvorkin:2014lea} 
  C.~Dvorkin, M.~Wyman, D.~H.~Rudd and W.~Hu,
  %``Neutrinos help reconcile Planck measurements with both Early and Local Universe,''
  arXiv:1403.8049 [astro-ph.CO].


\bibitem{Archidiacono:2014apa} 
M.~Archidiacono, N.~Fornengo, S.~Gariazzo, C.~Giunti, S.~Hannestad and M.~Laveder,  
 %``Light sterile neutrinos after BICEP-2,''
  arXiv:1404.1794 [astro-ph.CO].


\bibitem{Zhang:2014nta} 
  J.~-F.~Zhang, Y.~-H.~Li and X.~Zhang,
  %``Cosmological constraints on neutrinos after BICEP2,''
  arXiv:1404.3598 [astro-ph.CO].
  
\bibitem{DiValentino:2014zna}
  E.~Di Valentino, E.~Giusarma, M.~Lattanzi, A.~Melchiorri and O.~Mena,
  %``Axion cold dark matter: status after Planck and BICEP2,''
  arXiv:1405.1860 [astro-ph.CO].
  

\bibitem{Gerbino:2014eqa} 
  M.~Gerbino, A.~Marchini, L.~Pagano, L.~Salvati, E.~Di Valentino and A.~Melchiorri,
  %``Blue Gravity Waves from BICEP2 ?,''
  arXiv:1403.5732 [astro-ph.CO].
  

\bibitem{Cheng:2014bma} 
  C.~Cheng and Q.~-G.~Huang,
  %``The Tilt of Primordial Gravitational Waves Spectra from BICEP2,''
  arXiv:1403.5463 [astro-ph.CO].
  
\bibitem{Cheng:2014hba} 
  C.~Cheng and Q.~-G.~Huang,
  %``Probing the primordial Universe from the low-multipole CMB data,''
  arXiv:1405.0349 [astro-ph.CO].



\bibitem{Mortonson:2014bja} 
  M.~J.~Mortonson and U.~Seljak,
  %``A joint analysis of Planck and BICEP2 B modes including dust polarization uncertainty,''
  arXiv:1405.5857 [astro-ph.CO].
  %%CITATION = ARXIV:1405.5857;%%

  \bibitem{Dodelson:2014exa}
  S.~Dodelson,
  %``How much can we learn about the physics of inflation?,''
  Phys.\ Rev.\ Lett.\  {\bf 112} (2014) 191301
  [arXiv:1403.6310 [astro-ph.CO]].
  
\bibitem{Mukhanov:2013tua}
  V.~Mukhanov,
  %``Quantum Cosmological Perturbations: Predictions and Observations,''
  Eur.\ Phys.\ J.\ C {\bf 73}  2486 (2013). 
  [arXiv:1303.3925 [astro-ph.CO]].
  
  
\bibitem{Lyth:2009zz}
  D.~H.~Lyth and A.~R.~Liddle,
  %``The primordial density perturbation: Cosmology, inflation and the origin of structure,''
  Cambridge, UK: Cambridge Univ. Press (2009),  497 pp. 
  
  
\bibitem{Starobinsky:1980te}
  A.~A.~Starobinsky,
  %``A New Type of Isotropic Cosmological Models Without Singularity,''
  Phys.\ Lett.\ B {\bf 91} 99 (1980).
  
  \bibitem{Linde:1983gd} 
  A.~D.~Linde,
  %``Chaotic Inflation,''
  Phys.\ Lett.\ B {\bf 129}, 177 (1983).
  


\bibitem{Freese:1990rb}
  K.~Freese, J.~A.~Frieman and A.~V.~Olinto,
  %``Natural inflation with pseudo - Nambu-Goldstone bosons,''
  Phys.\ Rev.\ Lett.\  {\bf 65} 3233  (1990).
  
  
\bibitem{Adams:1992bn}
  F.~C.~Adams, J.~R.~Bond, K.~Freese, J.~A.~Frieman and A.~V.~Olinto,
  %``Natural inflation: Particle physics models, power law spectra for large scale structure, and constraints from COBE,''
  Phys.\ Rev.\ D {\bf 47}  426  (1993).   
  [hep-ph/9207245].
  
  
  \bibitem{CORE}
  F.~R.~Bouchet {\it et al.}  [{\sl COrE} Collaboration],
  %``COrE (Cosmic Origins Explorer) A White Paper,''
  arXiv:1102.2181 [astro-ph.CO].
  
  
\bibitem{pixie}
  A.~Kogut, D.~J.~Fixsen, D.~T.~Chuss, J.~Dotson, E.~Dwek, M.~Halpern, G.~F.~Hinshaw and S.~M.~Meyer {\it et al.},
  %``The Primordial Inflation Explorer (PIXIE): A Nulling Polarimeter for Cosmic Microwave Background Observations,''
  JCAP {\bf 1107}  025  (2011). 
  [arXiv:1105.2044 [astro-ph.CO]].
  





\bibitem{Mukhanov:2005sc} 
  V.~Mukhanov,
  ``Physical foundations of cosmology,''
  Cambridge, UK: Univ. Pr. (2005) 421 p. 
  



 
\bibitem{camb}
  A.~Lewis, A.~Challinor and A.~Lasenby,
  %``Efficient Computation of CMB anisotropies in closed FRW models,''
  Astrophys.\ J.\  {\bf 538}, 473 (2000)
  [arXiv:astro-ph/9911177].
  
  
\bibitem{Lewis:2002ah}
  A.~Lewis and S.~Bridle,
  %``Cosmological parameters from CMB and other data: a Monte-Carlo approach,''
  Phys.\ Rev.\  D {\bf 66}, 103511 (2002)
  [arXiv:astro-ph/0205436].
  
  
\bibitem{Ade:2013ktc} 
  P.~A.~R.~Ade {\it et al.}  [Planck Collaboration],
  %``Planck 2013 results. I. Overview of products and scientific results,''
  arXiv:1303.5062 [astro-ph.CO].
  
\bibitem{Planck:2013kta} 
  P.~A.~R.~Ade {\it et al.}  [Planck Collaboration],
  %``Planck 2013 results. XV. CMB power spectra and likelihood,''
  arXiv:1303.5075 [astro-ph.CO].
  
  \bibitem{Bennett:2012fp} 
  C.~L.~Bennett, D.~Larson, J.~L.~Weiland, N.~Jarosik, G.~Hinshaw, N.~Odegard, K.~M.~Smith and R.~S.~Hill {\it et al.},
  %``Nine-Year Wilkinson Microwave Anisotropy Probe (WMAP) Observations: Final Maps and Results,''
  arXiv:1212.5225 [astro-ph.CO].
  
\bibitem{Bezrukov:2007ep} 
F.~L.~Bezrukov and M.~Shaposhnikov,
  %``The Standard Model Higgs boson as the inflaton,''
  Phys.\ Lett.\ B {\bf 659}, 703 (2008)
  [arXiv:0710.3755 [hep-th]].
  
  
\bibitem{Bezrukov:2013fka} 
  F.~Bezrukov,
  %``The Higgs field as an inflaton,''
  Class.\ Quant.\ Grav.\  {\bf 30}, 214001 (2013)
  [arXiv:1307.0708 [hep-ph]].
  
  \bibitem{Ferrara:2014ima} 
  S.~Ferrara, A.~Kehagias and A.~Riotto,
  %``The Imaginary Starobinsky Model,''
  arXiv:1403.5531 [hep-th].
  
 
 \bibitem{Cook:2014dga} 
  J.~L.~Cook, L.~M.~Krauss, A.~J.~Long and S.~Sabharwal,
  %``Is Higgs Inflation Dead?,''
  arXiv:1403.4971 [astro-ph.CO].
  \bibitem{Nakayama:2014koa} 
   K.~Nakayama and F.~Takahashi,
  %``Higgs Chaotic Inflation and the Primordial B-mode Polarization Discovered by BICEP2,''
  arXiv:1403.4132 [hep-ph].
 \bibitem{Creminelli:2014oaa} 
 P.~Creminelli, D.~L\'opez Nacir, M.~Simonovi\'c, G.~Trevisan and M.~Zaldarriaga,
  %``$\phi^2$ or not $\phi^2$: Checking the Simplest Universe,''
  arXiv:1404.1065 [astro-ph.CO].
  
  
\bibitem{Okada:2014lxa} 
  N.~Okada, V.~N.~\c{S}eno\u{g}uz and Q.~Shafi,
  %``Simple Inflationary Models in Light of BICEP2: an Update,''
  arXiv:1403.6403 [hep-ph].
  
  
\bibitem{DiBari:2014oja} 
  P.~Di Bari, S.~F.~King, C.~Luhn, A.~Merle and A.~Schmidt-May,
  %``Radiative Inflation and Dark Energy RIDEs Again after BICEP2,''
  arXiv:1404.0009 [hep-ph].
  
  
\bibitem{Kaloper:2014zba} 
  N.~Kaloper and A.~Lawrence,
  %``Natural Chaotic Inflation and UV Sensitivity,''
  arXiv:1404.2912 [hep-th].
  
\bibitem{Creminelli:2014fca} 
  P.~Creminelli, D.~L\'opez Nacir, M.~Simonovi\'c, G.~Trevisan and M.~Zaldarriaga,
  %``$\phi^2$ Inflation at its Endpoint,''
  arXiv:1405.6264 [astro-ph.CO].
  
  
  
  \bibitem{Freese:2014nla} 
  K.~Freese and W.~H.~Kinney,
  %``Natural Inflation: Consistency with Cosmic Microwave Background Observations of Planck and BICEP2,''
  arXiv:1403.5277 [astro-ph.CO].
  
  
  \bibitem{Kohri:2014rja} 
  K.~Kohri, C.~S.~Lim and C.~-M.~Lin,
  %``Distinguishing between Extra Natural Inflation and Natural Inflation after BICEP2,''
  arXiv:1405.0772 [hep-ph].
 \bibitem{Amendola:2012ys}  
  L.~Amendola {\it et al.}  [Euclid Theory Working Group Collaboration],
  %``Cosmology and fundamental physics with the Euclid satellite,''
  Living Rev.\ Rel.\  {\bf 16}, 6 (2013)
  [arXiv:1206.1225 [astro-ph.CO]].
  %%CITATION = ARXIV:1206.1225;%%
  %109 citations counted in INSPIRE as of 21 May 2014
  

\end{thebibliography}
\end{document}